\documentclass{revtex4}
\usepackage{epsfig,amssymb,amsmath,epstopdf}
\begin{document}
\title{Charge Imbalance in a Layered Structure of High Temperature Superconductors}
\author{\hspace{1cm}Majed Nashaat~$^{1,3}$ }
\author{Yu. M. Shukrinov~$^{1,2}$}
\author{K. V. Kulikov~$^{1,2}$}
\author{R. Dawood~$^{1,3}$}
\author{Hussein El Samman$^{4}$}
\author{Th. M. El Sherbini,~$^{3}$}
\address{$^{1}$ BLTP, JINR, Dubna, Moscow Region, 141980, Russia \\
$^{2}$ Dubna International University of Nature, Society, and Man, Dubna, Moscow region, 141980 Russia \\
$^{3}$Department of Physics, Cairo University, Cairo, Egypt\\
$^{4}$Department of Physics, Faculty of Science, Menoufiya University, Egypt}

\begin{abstract}
Quasiparticle injection devices are considered as one of the candidates for the high temperature superconductor transistors, which can operate at liquid nitrogen temperatures. In these devices the nonequilibrium effects are created by injecting quasiparticle current into a stack of intrinsic Josephson junctions (IJJs). These effects lead to an occurrence of a shift of the condensate chemical potential and a difference in the distribution between the electron-like and the hole-like quasiparticles that causes a charge imbalance. Such effects are observed in many experiments for both bulk and layered superconductors. In this paper, we study non-stationary nonequilibrium charge imbalance effect due to current injection in a stack of IJJs. We investigate the effect of the charge imbalance on the current-voltage characteristic (IVC) and the time dependence of the voltage and the quasiparticle potential in a stack of five IJJs.\\ \\
\textbf{Keywords:} Charge imbalance, Josephson junctions.
\\ \textbf{Accepted to be published in Egyptian Journal of Physics no:2 Vol 2015 }

\end{abstract}

\maketitle
\section{Introduction}
In strongly anisotropic layered high temperature superconductors such as ($Bi_2 Sr_2 Ca Cu_2 O_8$), the superconducting layers $CuO_2$ (S-layers) together with intermediate dielectric layers ($BiO, SrO$) form the system of intrinsic Josephson junctions (IJJs)\cite{Kleiner}. The thickness of the S-layers in a high-temperature superconductor is comparable to the screening length of the electric charge and charge relaxation length. For this reason, the screening of the charge in an individual S-layer is incomplete and the electric field induced in an individual Josephson junction penetrates to the neighboring junctions. The electric neutrality of the S-layers is dynamically broken. When an external electric current flows through the IJJs, the superconducting layers are in a nonequilibrium state because of the injection of quasiparticles and Cooper pairs\cite{Dmitry}. Since an uncompensated electric charge exists in the junctions, an additional current between superconducting layers should be taken into account. This contribution to the quasiparticle current owing to the difference between generalized scalar potentials is called diffusion current \cite{Keller,Artemenko,Shukrinov}. In the absence of the complete screening of the charge in the S-layer, the Josephson relation for the gauge invariant phase difference between S-layers is generalized. In contrast to the equilibrium case, the derivative of the phase difference in the l-th junction depends now not only on the voltage $V_l  ≡ V_{(l,l-1)}$ in this junction but also on the voltages in the neighboring ((l -1)th and (l + 1)th) Josephson junctions. 

The dynamics of a stack of IJJs in HTS are theoretically investigated previously with the quasi-neutrality breakdown effect \cite{Koyama} and the quasiparticle charge imbalance effect \cite{Dmitry,Keller,Rother,DARyndyk,Shukrinov2} . In Ref.\cite{Dmitry}, two effects are observed experimentally for the stationary case (i.e. no displacement current) which result from the charge imbalance in the S-layers. The first one is the shifting of the Shapiro step from the canonical value $\hbar f/(2e)$, where f is the frequency of radiation, $\hbar$ is Planck’s constant and e is the electric charge.  The second effect occurs when two mesas structures are close to each other on the same base crystal and the influence of the current through one mesa on the voltage drops across the other mesa is measured.

The charge imbalance effect is essential if the layer thickness $d_s$ is smaller than the characteristic length of the nonequilibrium relaxation $l_e$ \cite{Masahiko,Tinkham,NArtemenko}. In the layered superconductors with IJJs nonequilibrium effects become important since the effective layer thickness is small (3 $\sim$ 10 $\AA{}$) and a typical value of $l_e$ is about 1 $\mu m$, so that $d_s$ $<< $ $l_e$ \cite{Ryndyk3}.

In this paper we present the results of the investigation of the time dependence of the voltage in each Josephson junction and the quasiparticle potential in each S-layer in the stack of IJJs in the presence of non-stationary nonequilibrium charge imbalance effect. We use the capacitively coupled Josephson junctions model (CCJJ+DC) including the charge imbalance effect (CIB). 
\section{Model}
Let us consider a system of N+1 S-layers in anisotropic high-Tc superconductor shown in Fig. 1. The thickness of the superconducting and the insulating layers are denoted by $d_{s}^{l}$ and $d$, respectively. At the edges of the stack the effective thickness of the S-layer can be extended due to the proximity effect into the attached metals. The thicknesses of the 0th and Nth S-layers are denoted by $d_{s}^{0}$ and $d_{s}^{N}$,  respectively. All other S-layers and the insulating layers in between are of the same thickness. The N+1 S-layers are characterized by the order parameter $\Delta_{l} (t)$=$exp$ $[i \chi_{l} (t)]$ with the time dependent phase  $\chi_{l} (t)$ and form N Josephson junctions \cite{Kleiner}. The gauge invariant phase difference and the voltage of each IJJ are represented by $\phi_{(l,l+1)}\equiv\phi_{l}$ and $V_{l}$, respectively, while the tunneling current is represented by $J_{0}^{qp}=J_{N}^{qp}=J$ for the first and the last S-layers and by $J_{(l,l+1)}$ for the middle layers (see Fig.1). The quasiparticle potential of each S-layer is represented by $\psi_{l}$.

\begin{figure}[h!]
	\centering
	\includegraphics[width=.7\linewidth, angle =0]{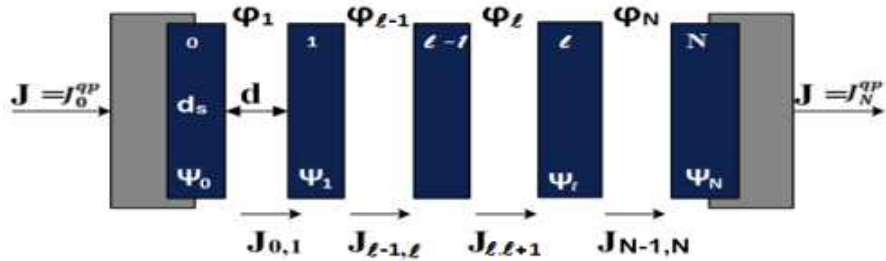}
	\caption{ (Color online) Schematic diagram of a system of IJJs in HTS. Superconducting layers, which are numbered by l running from 0 to N, form a system of IJJs with phase differences  $\phi _{l}=\chi_{l}-\chi_{(l-1)}$; $d_{s}$ and d are the thicknesses of the superconducting and dielectric layers, respectively. The charge imbalance is represented by the quasiparticle potential $\psi_{l}$ while the current through the junctions is represented by $J_{(l,l+1)}$. The first and the last S-layer have  $J_{(0)}$=J=$J_{(N+1)}$.}
	\label{fig1}
\end{figure}

The system of equations describing this electric scheme includes: the generalized Josephson relation with charge imbalance and the total current given by the CIB model, in addition to the kinetic equations for the quasiparticle potential in the S-layers \cite{Rother}

\begin{eqnarray}
\dot{\varphi}_{1} (t)&=& (1+2\alpha) v_{1} - \alpha (v_{l-1} + v_{1+1}) +\dfrac{\psi_{1}-\psi_{l-1}}{\beta}\\
\dot{v}_{l} (t)&=&   \frac{J}{J_{c}} -\sin\varphi_{l}-\beta\dot{\varphi}_{l}+\psi_{l}-\psi_{l-1}\\
\zeta_{l} \dot{\psi}_{l}&=& \eta_{l} \left( J_{l}^{qp} - J_{l+1}^{qp} \right) -\psi_{l}
\label{eq:systemofeq}
\end{eqnarray}

where $v_l$ is the voltage, normalized to $V_{0}=\hbar \omega_{p}/2e$,  $\omega_{p}^{2}=2eJ_{c}/\hbar C$ is the plasma frequency, $J_{c}$ is the critical current of the junction and C is the capacitance of the junctions,  $\alpha$ is the coupling parameter between IJJs, $\beta=1/R  (\hbar/(2eJ_{c} C)^{1/2})$ is the dissipation parameter (if $\beta$ $\textless \textless$ 1 we have under-damped case, while for $\beta$ $\textgreater \textgreater $ 1 we have over-damped case \cite{Koyama}), R is the junction resistance, $\tau = \omega_{p} t$ is the normalized time, $\Psi=\psi/RJ_{c}$ is the normalized quasi-particle potential, $\xi_{l}$  is the normalized quasi-particle relaxation time,  $\eta_{l}$ = $4π r_{D}^{2}/d_{s}^{l}$  is the non-equilibrium parameter, $r_{D}$ is the penetration depth of the electric field and  $J_{l}^{qp}$ is the quasi-particle current of each S-layer, dot means the time. An estimation of these parameters can be found in Ref.\cite{Ryndyk3,Krasnov,Winkler,Yurgens,Krasnov2}. Eq.(1) shows that the phase of each junction depends on the voltage of the neighboring junctions and the difference between the quasi-particle potential of the neighboring S-layers. Eq.(2) represents the total current in the junction in the normalized form. Current through the IJJ consists of the displacement current (C dV/dt), the supercurrent ($J_{c} \sin \phi$) and the quasi-particle current $((\hbar \dot{\varphi})/2eR+  ((\Psi_(l-1)-\Psi_l))/R)$ as found by the CIB and CCJJ+DC models \cite{Shukrinov2,Shukrinov3}.
Eq.(3) represents the kinetic equations for charge imbalance potentials in the CIB model. The first term is proportional to the tunneling of the quasiparticles through the IJJs, while the second term represents the relaxation of the quasiparticles in each S-layer. We solve this system of equations numerically using the fourth order Runge-Kutta method.

\section{Results}

In this section we present the IVC and time dependence of the voltage and the quasiparticle potential in a stack of IJJs. We consider a system of five IJJs with $\beta$=0.2 (under-damped case, where the IV characteristic has a hysteresis region \cite{Charles}). 
Fig. 2 demonstrates the IVC of stack at $\beta$=0.2 and $\alpha$=0. Fig.2 (a) shows the IVC of the system without charge imbalance effect ($\eta$=0,$\xi$=0). In Fig.2 (b) the one loop IVC of the stack with $\eta_{l}$=0.4, $\xi_{l}$  =0.1 is shown. The IVC demonstrates a step structure near $I_{c}$ and in the hysteresis region. These steps correspond to the different branches as shown in the inset of Fig. 2 (b).
\begin{figure}[h!]
	\begin{minipage}{.5\textwidth}	
		\centering
		\includegraphics[width=.7\linewidth, angle =-90]{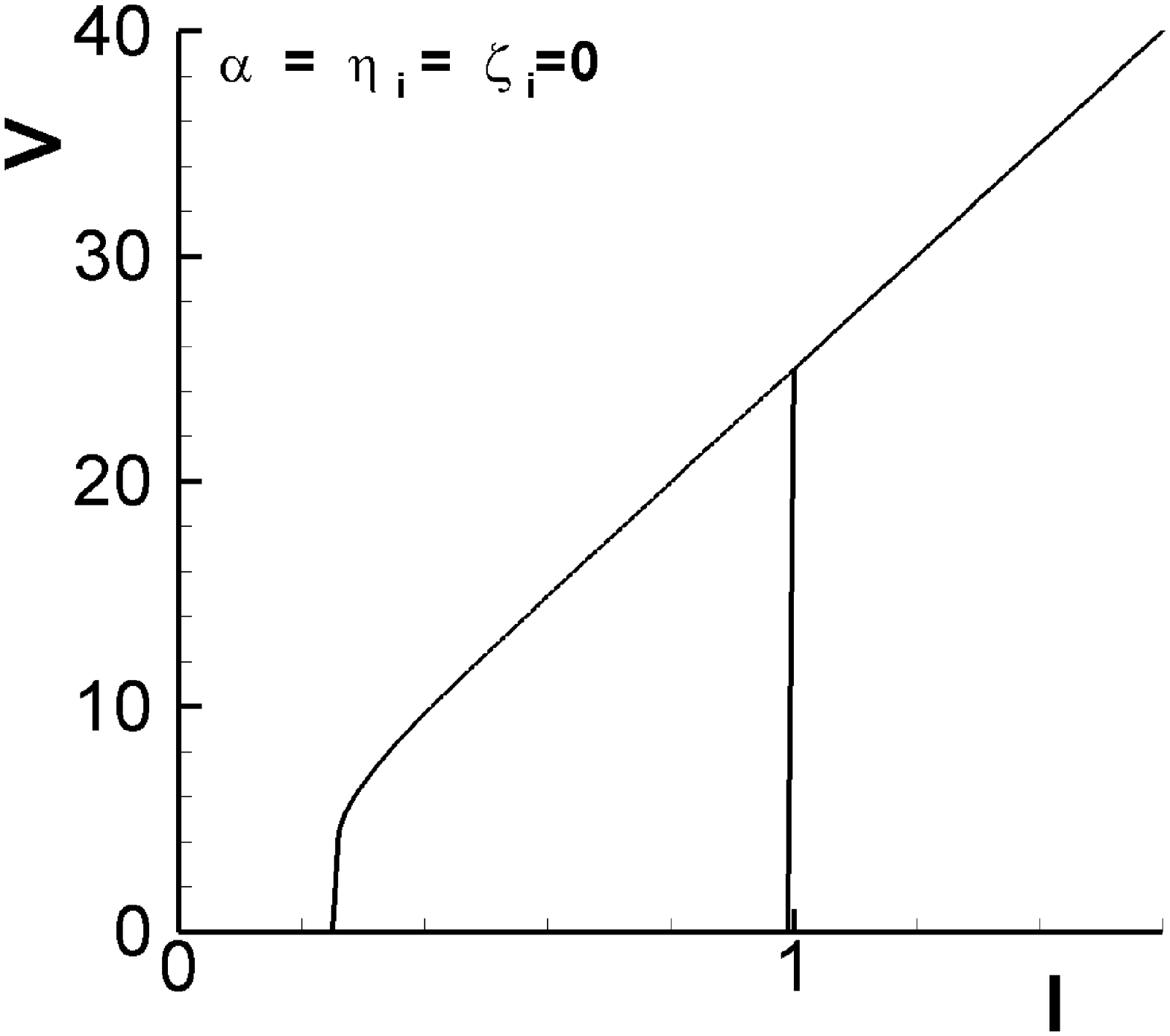}\\(a)
	\end{minipage}%	
	\begin{minipage}{.5\textwidth}		
		\includegraphics[width=.7\linewidth, angle =-90]{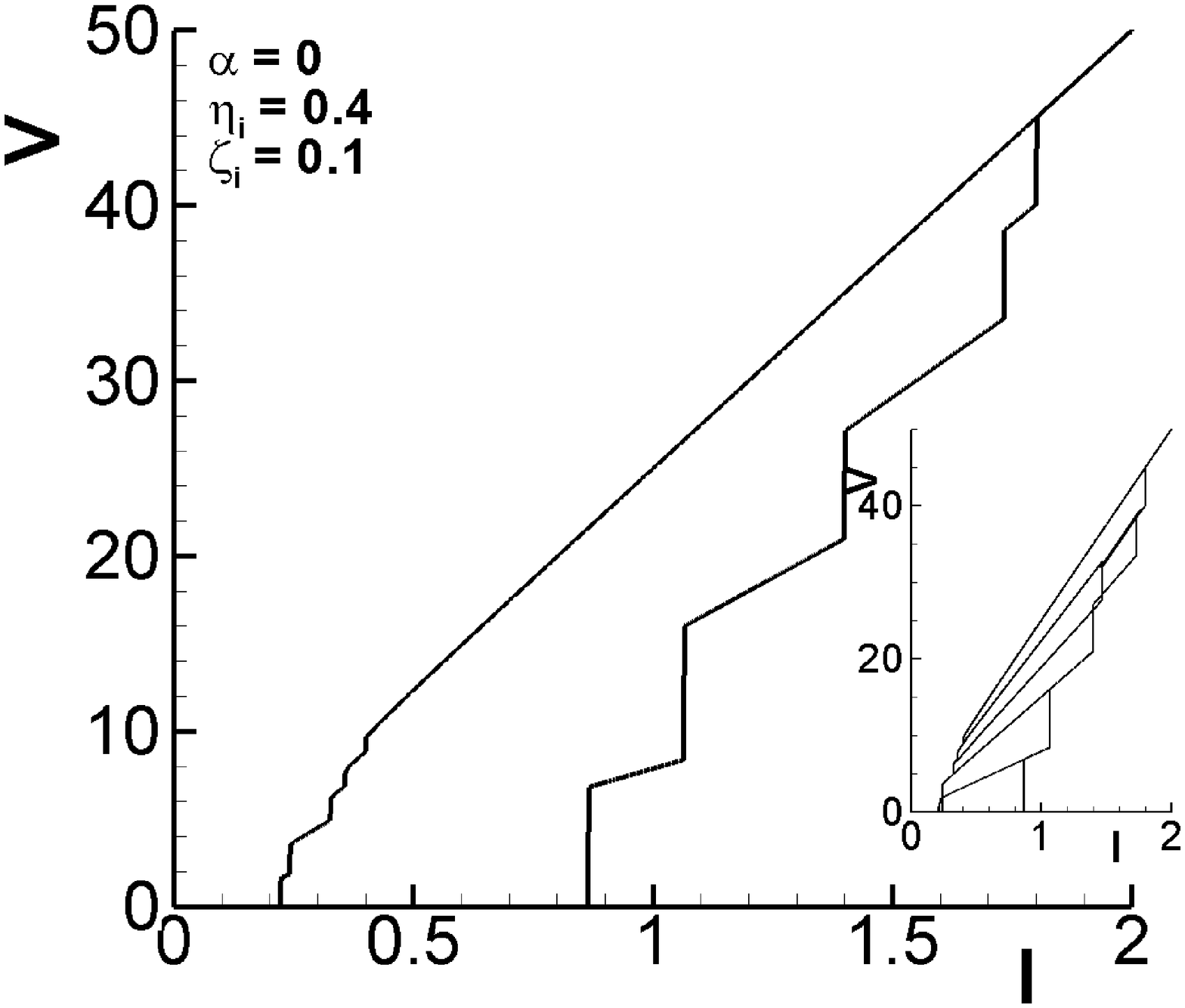} \\
		(b)
	\end{minipage}%	
	\caption {One loop IV characteristic of a stack of 5 IJJs with periodic boundary condition ($v_{N}=v_{0}$, $v_{(N+1)}=v_{1}$, and $\psi_{N}=\psi_{0}$, $\psi_{N+1}=\psi_{1}$) (a) without the charge imbalance effect. (b) with the charge imbalance effect ($\eta_{l}$ = 0.4, $\xi_{l}$ =0.1 ). The inset shows the branched IV characteristic. If $\alpha$= 0 the IJJs are not completely decoupled from each other(see Eq.(1)).} 

\end{figure} %

If one of IJJs switches into resistive state, the nonequilibrium quasiparticle distribution is induced in the neighboring junctions and give rise to a charge imbalance. The presence of a charge imbalance in the S-layers leads to the appearance of a quasiparticle current through the neighboring junctions. A decrease in the supercurrent through these junctions occurs. In order to switch these junctions to the resistive state a larger external current is needed. Such process is known as “current effect” \cite{Dmitry,Ryndyk3}. The branches in the IV characteristic correspond to different number of the IJJs in the rotating (R-state) and in the oscillating state (O-state). The time average $\textless d\varphi/dt\textgreater$ is constant and $\textless \sin\varphi\textgreater$ is zero for a junction in the rotating state, while for the oscillating state, the time average $\textless d\varphi/dt\textgreater$  is zero and $\textless \sin\varphi\textgreater$ is constant \cite{Shukrinov3,Hideki}.\\

\begin{figure}[h!]
	\centering
	\includegraphics[width=.3\linewidth, angle =-90]{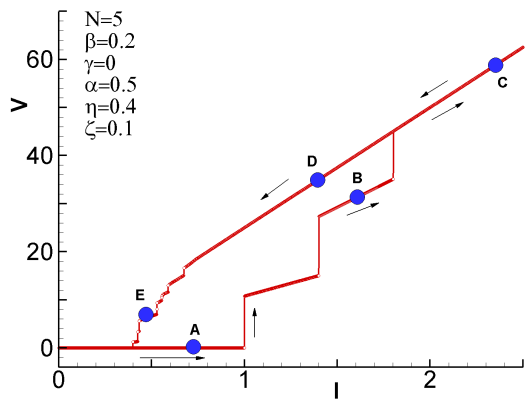}
	\caption{ One loop IV characteristic of a stack of 5 IJJs, the arrows show the direction of the current while the circles represent the points at which we investigated the time dependence for voltage and quasiparticle potentials.}
\end{figure}

Fig. 3 demonstrates the IVC of 5 IJJs at $\beta$=0.2 and $\alpha$=0.5, $\eta_{l}$=0.4 and $\xi_{l}$=0.1. Arrows indicate the direction of current sweeping. We investigate time dependence of the voltage in each IJJs and the quasi-particle potential in each S-layer in different regions of the IVC. \\

The first point A of Fig.3 corresponds to the zero voltage state. In this region all the IJJs are in the O-state and the current is carried by the Cooper pairs only. The second point B corresponds to the state in which first, third and fifth IJJs are in the R-state (R(1,3,5)), while the second and fourth junctions are in the O-state. Points C and D lies in the outermost branch (i.e. state R(1,2,3,4,5)) and point E correspond to the state R(1,4).\\

The time dependences of the voltage for the points B, C, D and E are shown in Fig. 4. Fig. 4 (a) demonstrates time dependence of the voltage for each IJJ at point B.  The voltage of the IJJ in the state R(1,3,5) is oscillating (see dashed thick, long dashed thin and dashed thin curve in Fig. 4 (a) respectively). The voltage for the IJJs in O-state is a straight line in this scale (see solid thick and solid thin curve). However, in the inset we can see small ripples in the voltage for these junctions, as a result of the coupling between IJJs and the charge imbalance on the S-layers. These ripples depend on the difference in the quasiparticle potential on the neighboring S-layers (see Eq.(1)).

\begin{figure}[h!]
	\centering
	\includegraphics[width=.6\linewidth, angle =0]{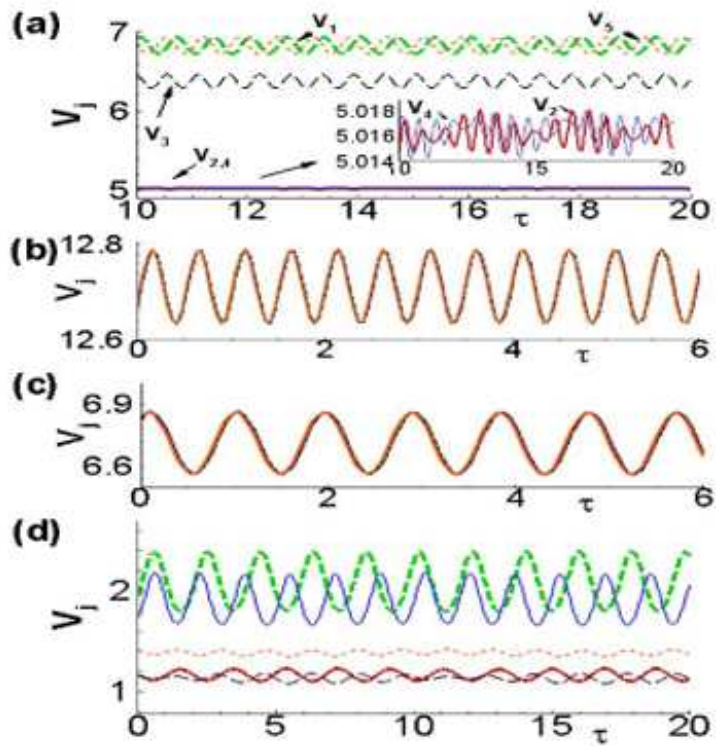}
	\caption{ (color online) Time dependence of voltage in each IJJs in the stack: (a) at point B, (b) at point C, (c) at point D, (d) at point E. $v_{l}$ is the voltage in the l-th IJJ. Dashed thick line (green online) related to first IJJ, solid thick line (red online) related to second IJJ, long dashed thin line (black online) related to third IJJ, solid thin line (blue online) related to fourth IJJ, dashed thin line (orange online) related to fifth IJJ.}
\end{figure}

In Fig. 4 (b,c) the time dependences of voltage at points C and D are shown respectively. The voltages of all junctions oscillate with the same amplitude and frequency. The frequency of voltage oscillations at point D (which corresponds to the hysteresis region) is less than the frequency at point C, because these points correspond to different voltages in the IV characteristic (see Fig. 3). Fig. 4 (d) presents the time dependence of voltage at points E. The voltage of the IJJs in R(1,4) oscillate harmonically (dashed and solid curve, respectively). The IJJs in O-state demonstrate small ripples in the voltage due to the charge imbalance in the neighboring S-layers which lie between resistive and superconducting junctions. \\

Fig.5 demonstrates time dependence of the quasiparticle potential in S-layers in the stack of IJJs. Fig. 5 (a-1,2,3) shows the quasiparticle potential at point B for the first, second and third S-layers respectively. The potential of the first S-layer $\psi_{0}$ oscillates harmonically and its average value is  $\textless \psi_{0}\textgreater$ =0 (see Fig. 5 (a-1)), while $\psi_{1,2}$ have modulated oscillations and their average value have different signs $\textless \psi_{1,2}\textgreater$ $\neq$0 (see Fig. 5 (a-2,3)). 

Fig. 5 (b-1,2,3) and (c-1,2,3) show time dependence of the quasiparticle potential at point C and D for the first, second and third S-layers respectively. These points correspond to the outermost branch (all IJJ are in R-state) and all $\psi_{l}$ here oscillate harmonically around zero value. The amplitude of oscillations is maximal at the edges (see Fig. 5 (b-1,c-1)) and decreases to the middle (see Fig. 5 (b-3,c-3)). The average value of $\textless \psi_{l}-\psi_{l-1}\textgreater$ =0, as a result there is no influence of charge imbalance on the IVC in this case.

\begin{figure}[h!]
	\centering
	\includegraphics[width=.6\linewidth, angle =0]{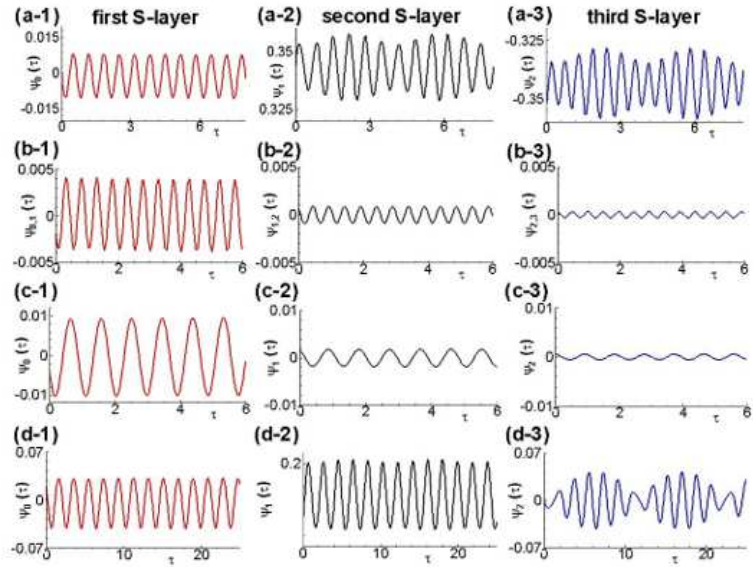}
	\caption{ (color online) . Time dependence of the quasiparticle potential in S-layers in the stack for different regions in the IV characteristic shown in Fig.3. Point B corresponds to the R(1,3,5), points C and D correspond to all the IJJs in the R-state outside and inside the hysteresis region, respectively, and point E corresponds to the R(1,4). }
\end{figure}

Fig. 5 (d-1,2,3) demonstrates time dependence of the quasiparticle potential at point E for the first, second and third S-layers respectively. The potential of the first S-layer $\psi_{0}$ oscillate harmonically and its average value is $\textless \psi_{0}\textgreater$ =0 (see Fig. 5 (d-1)). Fig. 5 (d-2) shows that the potential of the second S-layer  $\psi_{1}$ oscillates with a small modulation and its average value is $\textless \psi_{1}\textgreater$ $\neq$0. The quasiparticle potential $\psi_{2}$ has a modulated oscillations and  $\textless \psi_{0}\textgreater$ =0 (see Fig. 5 (d-3)). As it can be seen from Fig. 5 (a-1,2,3) and (d-1,2,3), the amplitude of the quasiparticle potential is large for the S-layer which is placed between resistive and superconducting junction. This behavior is consistent with the results for the quasiparticle potential in Ref.\cite{Rother} for the stationary case.

\section{Conclusions}

We have investigated the effect of charge imbalance on the IV characteristic in a stack of five IJJs. In the case of nonstationary charge imbalance in the stack of IJJs, we have found that not all the junctions are collectively switched to the resistive state, which is due to the current effect. We have simulated the branch structure of the IV characteristic and have investigated the time dependence of voltage for each IJJs and the quasiparticle potential for each S-layer at different points of the IV characteristic. Consistent with other authors we have found that the charge imbalance becomes stronger when the superconducting layer is placed between resistive and superconducting junctions.

\textbf{Acknowledgement}
We acknowledge the Joint Institute for Nuclear Research (JINR)-EGYPT collaboration and thank T. Hussein, Kh. Hegab, D.V. Kamanin, and A. Elithi for support of this paper.

\end{document}